\documentclass[preprint,12pt]{elsarticle}

\usepackage[english]{babel}

\usepackage[utf8]{inputenc}
\usepackage[T1]{fontenc}

\usepackage{verbatim}

\usepackage{orcidlink}

\PassOptionsToPackage{hyphens}{url}

\usepackage{url}                
\usepackage{graphicx}           

\usepackage[
    binary-units,
    range-units=single,
    list-units=single,
    per-mode=symbol,
    per-symbol=p,
    detect-all
]{siunitx}                      
\usepackage{relsize}            
\usepackage{xspace}             
\usepackage{calc}               
\usepackage{microtype}          
\usepackage[inline]{enumitem}   
\usepackage{nth}                
\usepackage{import}             
\usepackage[noend,figure,vlined]{algorithm2e} 
\usepackage{footnote}           

\usepackage{keycommand}         

\usepackage{hyphenat}           

\usepackage{bibentry}           

\usepackage{flafter}

\usepackage{dpfloat}

\usepackage{array}              
\usepackage{longtable}          
\usepackage{multirow}           
\usepackage{makecell}           
\usepackage{tabularx}           
\usepackage{booktabs}           

\usepackage{pgffor}             
\usepackage{adjustbox}          

\usepackage{circledsteps}       
\usepackage{nicefrac}

\usepackage{amsmath}            
\usepackage{amssymb}            
\usepackage{amsfonts}           
\usepackage{amsthm}             
\usepackage{mathtools}          

\usepackage{caption}
\usepackage{subcaption}

\usepackage{hyperref}

\usepackage[capitalise,nameinlink,noabbrev]{cleveref}

\usepackage[
    nomain,             
    acronym
]{glossaries-extra}

\newignoredglossary{ignored-glossary}

\usepackage{ifdraft}

\usepackage[colorinlistoftodos,prependcaption,textsize=tiny,obeyFinal]{todonotes} 

\glsenableentrycount
\glssetcategoryattribute{abbreviation}{entrycount}{1}

\setdescription{topsep=0pt,leftmargin=0pt,labelindent=\parindent}

\protected\def\nmicro{\text{\ensuremath{\mu}}}
\glsenableentrycount
\let\gls\cgls
\let\glspl\cglspl
\let\Gls\cGls
\let\Glspl\cGlspl

\glspatchtabularx

\glsdisablehyper

\pgfkeys{/csteps/inner color=white}
\pgfkeys{/csteps/inner xsep=7pt}
\pgfkeys{/csteps/outer color=black}
\pgfkeys{/csteps/fill color=black}
\newcommand\circleref[1]{{\;\smaller\Circled[]{\textbf{#1}}}}

\definecolor{darkgreen}{RGB}{0, 120, 0}

\DeclareSIUnit{\nothing}{\relax}

\makeatletter
\newcommand{\etal}{et al\@ifnextchar{.}{\xspace}{.\@\xspace}}
\makeatother

\newcommand\quotes[1]{``{#1}''}

\newcolumntype{L}{X}
\newcolumntype{R}{>{\raggedleft\arraybackslash}X}
\newcolumntype{C}{>{\centering\arraybackslash}X}

\newcolumntype{K}[1]{>{\raggedright\arraybackslash}m{#1}}
\newcolumntype{M}[1]{>{\centering\arraybackslash}m{#1}}
\newcolumntype{H}[1]{>{\raggedleft\arraybackslash}m{#1}}

\newcommand{\ccode}[1]{\mbox{\texttt{#1}}}

\providecommand{\rebuttal}[1]{#1}
\providecommand{\rebuttaltwo}[1]{#1}

\newcommand\mhz{\mega\hertz}

\glsenableentrycount

\newcommand\xavierboard{NVIDIA Jetson AGX Xavier\xspace}

\DeclareSIUnit{\pkt}{p}
\DeclareSIUnit{\b}{b}
\DeclareSIUnit{\bits}{bits}
\DeclareSIUnit{\bytew}{byte}
\DeclareSIUnit{\bytes}{bytes}
\DeclareSIUnit{\packets}{packets}

\mathchardef\mhyphen="2D

\newabbreviation[longplural={Network Points of Presence}]{npop}{N-PoP}{Network Point of Presence}
\newabbreviation[longplural={system-on-chips}]{soc}{SoC}{system-on-a-chip}
\newabbreviation[shortplural={CAPEX}]{capex}{CAPEX}{Capital Expenditure}
\newabbreviation[shortplural={IDS}]{ids}{IDS}{Intrusion Detection System}
\newabbreviation[shortplural={OPEX}]{opex}{OPEX}{Operating Expense}

\newabbreviation{3gpp}{3GPP}{3rd Generation Partnership Project}
\newabbreviation{ac}{AC}{access control}
\newabbreviation{adas}{ADAS}{Advanced Driver-Assistance System}
\newabbreviation{adc}{ADC}{analog-to-digital converter}
\newabbreviation{alu}{ALU}{Arithmetic-Logic Unit}
\newabbreviation{amf}{AMF}{Access and Mobility Management Function}
\newabbreviation{api}{API}{application-programming interface}
\newabbreviation{atet}{AT\&T}{American Telephone \& Telegraph}
\newabbreviation{awsecs}{ECS}{Elastic Container Service}
\newabbreviation{bkl}{BKL}{Big Kernel Lock}
\newabbreviation{bpu}{BPU}{Branch Prediction Unit}
\newabbreviation{bt}{BT}{British Telecom}
\newabbreviation{cbs}{CBS}{Constant Bandwidth Server}
\newabbreviation{cfs}{CFS}{Completely Fair Scheduler}
\newabbreviation{cmos}{CMOS}{Complementary Metal–Oxide–Semiconductor}
\newabbreviation{cn}{CN}{Core Network}
\newabbreviation{cots}{COTS}{Commercial Off-the-Shelf}
\newabbreviation{csr}{CSR}{control and status register}
\newabbreviation{cu}{CU}{Central Unit}
\newabbreviation{cu/du}{CU/DU}{Central Unit/Distributed Unit}
\newabbreviation{cupti}{CUPTI}{Cuda Performance Tools Interface}
\newabbreviation{dag}{DAG}{Directed Acyclic Graph}
\newabbreviation{ddio}{DDIO}{Data Direct I/O Technology}
\newabbreviation{dlp}{DLP}{Data-Level Parallelism}
\newabbreviation{dpdk-ans}{DPDK-ANS}{DPDK Accelerated Network Stack}
\newabbreviation{dpdk}{DPDK}{Data Plane Development Kit}
\newabbreviation{dpi}{DPI}{deep packet inspection}
\newabbreviation{dpm}{DPM}{Dynamic Power Management}
\newabbreviation{dpr}{DPR}{Dynamic Partial Reconfiguration}
\newabbreviation{dprio}{DP}{Dynamic Priority}
\newabbreviation{dt}{DT}{Deutsche Telekom}
\newabbreviation{du}{DU}{Distributed Unit}
\newabbreviation{dvfs}{DVFS}{Dynamic Voltage and Frequency Scaling}
\newabbreviation{dvs}{DVS}{vSphere Distributed Switch}
\newabbreviation{eal}{EAL}{Environment Abstraction Layer}
\newabbreviation{edd}{EDD}{Event-Driven Delay-Induced}
\newabbreviation{edf}{EDF}{Earliest Deadline First}
\newabbreviation{eetb}{EETB}{estimated execution time bound}
\newabbreviation{egid}{EGID}{effective group ID}
\newabbreviation{embb}{eMBB}{enhanced Mobile Broadband}
\newabbreviation{enb}{eNB}{Evolved NodeB}
\newabbreviation{epc}{EPC}{Evolved Packet Core}
\newabbreviation{etsi}{ETSI}{European Telecommunications Standards Institute}
\newabbreviation{euid}{EUID}{effective user ID}
\newabbreviation{f1ap}{F1AP}{F1 Application Protocol}
\newabbreviation{f1c}{F1-C}{F1 Control Plane Interface}
\newabbreviation{f1oip}{F1oIP}{F1-over-IP}
\newabbreviation{f1u}{F1-U}{F1 User Plane Interface}
\newabbreviation{fdio}{FD.io}{Fast Data Project}
\newabbreviation{fifo}{FIFO}{First In First Out}
\newabbreviation{first}{FIRST}{Flexible Integrate Real-time Scheduling Technologies}
\newabbreviation{fiveg}{5G}{5th Generation}
\newabbreviation{fpga}{FPGA}{Field Programmable Gate Array}
\newabbreviation{fprio}{FP}{Fixed Priority}
\newabbreviation{fpu}{FPU}{Floating-Point Unit}
\newabbreviation{gedf}{G-EDF}{Global Earliest Deadline First}
\newabbreviation{gnb}{gNB}{Next Generation NodeB}
\newabbreviation{gtp}{GTP}{GPRS Tunneling Protocol}
\newabbreviation{gui}{GUI}{graphical user interface}
\newabbreviation{hpc}{HPC}{High Performance Computing}
\newabbreviation{hss}{HSS}{Home Subscriber Service}
\newabbreviation{i2c}{I2C}{Inter Integrated Circuit}
\newabbreviation{iaas}{IaaS}{Infrastructure as a Service}
\newabbreviation{ict}{ICT}{Information and Communications Technology}
\newabbreviation{ilp}{ILP}{Integer Linear Programming}
\newabbreviation{iommu}{IOMMU}{Input-Output Memory Management Unit}
\newabbreviation{iot}{IoT}{Internet of Things}
\newabbreviation{ipc}{IPC}{Inter Process Communication}
\newabbreviation{isa}{ISA}{Instruction Set Architecture}
\newabbreviation{isg}{ISG}{Industry Specification Group}
\newabbreviation{itu}{ITU}{International Mobile Telecommunications}
\newabbreviation{itur}{ITU-R}{ITU Radiocommunication Sector}
\newabbreviation{let}{LET}{Logical Execution Time}
\newabbreviation{libos}{LibOS}{Library Operating System}
\newabbreviation{libosnuse}{LibOS-NUSE}{LibOS Network Stack in Userspace} 
\newabbreviation{llc}{LLC}{Last Level Cache}
\newabbreviation{lls}{LLS}{Linear Least Squares}
\newabbreviation{lte}{LTE}{Long Term Evolution}
\newabbreviation{lut}{LUT}{Lookup Table}
\newabbreviation{lxc}{LXC}{Linux Containers}
\newabbreviation{mac}{MAC}{Media Access Control}
\newabbreviation{mano}{MANO}{Management and Orchestration}
\newabbreviation{mde}{MDE}{Model-Driven Engineering}
\newabbreviation{milp}{MILP}{Mixed-Integer Linear Programming}
\newabbreviation{miqcp}{MIQCP}{Mixed-Integer Quadratic Constraint Programming}
\newabbreviation{mme}{MME}{Mobility Management Entity}
\newabbreviation{mmtc}{mMTC}{massive Machine-Type Communications}
\newabbreviation{mp2soc}{MP2SoC}{Massively Parallel Multi-Processors System-on-Chip}
\newabbreviation{napi}{NAPI}{Linux New API}
\newabbreviation{nas}{NAS}{Network Attached Storage}
\newabbreviation{nfv}{NFV}{Network Function Virtualization}
\newabbreviation{nfvi}{NFVI}{Network Function Virtualization Infrastructure}
\newabbreviation{nfvo}{NFVO}{NFV Orchestrator}
\newabbreviation{ng}{NG}{next-generation}
\newabbreviation{ngc}{NGC}{Next-Generation Core}
\newabbreviation{ngran}{NG-RAN}{Next-Generation RAN} 
\newabbreviation{nic}{NIC}{Network Interface Controller}
\newabbreviation{nnls}{NNLS}{Non-Negative Least Squares}
\newabbreviation{nsa}{NSA}{Non-SA} 
\newabbreviation{numa}{NUMA}{Non-Uniform Memory Access}
\newabbreviation{oai}{OAI}{OpenAirInterface}
\newabbreviation{oran}{O-RAN}{Open RAN} 
\newabbreviation{osa}{OSA}{OAI Software Alliance} 
\newabbreviation{ovs}{OVS}{Open vSwitch}
\newabbreviation{paae}{PAAE}{Percentage Average Absolute Error}
\newabbreviation{paas}{PaaS}{Platform as a Service}
\newabbreviation{pcb}{PCB}{printed circuit board}
\newabbreviation{pcc}{PCC}{Pearson Correlation Coefficient}
\newabbreviation{pcp}{PCP}{Priority Ceiling Protocol}
\newabbreviation{pe}{PE}{Processing Element}
\newabbreviation{pedf}{P-EDF}{Partitioned Earliest Deadline First}
\newabbreviation{pf}{PF}{Physical Function}
\newabbreviation{phy}{PHY}{physical layer}
\newabbreviation{pid}{PID}{process ID}
\newabbreviation{pip}{PIP}{Priority Inheritance Protocol}
\newabbreviation{pl}{PL}{Programmable Logic}
\newabbreviation{pm}{PM}{Poll Mode Driver}
\newabbreviation{pmc}{PMC}{Performance Monitoring Counter}
\newabbreviation{pmd}{PMD}{Poll Mode Driver}
\newabbreviation{pmu}{PMU}{Performance Monitor Unit}
\newabbreviation{pnf}{PNF}{Physical Network Function}
\newabbreviation{ps}{PS}{Processing System}
\newabbreviation{psu}{PSU}{Power Supply Unit}
\newabbreviation{pu}{PU}{processing unit}
\newabbreviation{qos}{QoS}{Quality of Service}
\newabbreviation{ran}{RAN}{Radio Access Network}
\newabbreviation{rapl}{RAPL}{Running Average Power Limit}
\newabbreviation{rdma}{RDMA}{Remote Direct Memory Access}
\newabbreviation{rm}{RM}{Rate Monotonic}
\newabbreviation{rr}{RR}{Round Robin}
\newabbreviation{rrc}{RRC}{Radio Resource Control}
\newabbreviation{rru}{RRU}{Remote Radio Unit}
\newabbreviation{rt}{RT}{real-time}
\newabbreviation{rtkit}{RTKit}{RealtimeKit}
\newabbreviation{rtt}{RTT}{round-trip time}
\newabbreviation{ru}{RU}{Remote Unit}
\newabbreviation{sa}{SA}{Stand Alone}
\newabbreviation{san}{SAN}{Storage Area Network}
\newabbreviation{sba}{SBA}{service-based architecture}
\newabbreviation{sctp}{SCTP}{Stream Control Transmission Protocol}
\newabbreviation{sdn}{SDN}{Software-Defined Networking}
\newabbreviation{simd}{SIMD}{Single Instruction, Multiple Data}
\newabbreviation{sm}{SM}{Streaming Multiprocessor}
\newabbreviation{smf}{SMF}{Session Management Function}
\newabbreviation{spgw}{SPGW}{Serving Gateway/PDN Gateway}
\newabbreviation{sriov}{SR-IOV}{Single-Root I/O Virtualization}
\newabbreviation{srp}{SRP}{Stack Resource Policy}
\newabbreviation{sse}{SSE}{Streaming SIMD Extensions}
\newabbreviation{stream}{STREAM}{Simulation Tool for Energy Efficient Real Time Scheduling and Analysis}
\newabbreviation{svr}{SVR}{Support Vector Regression}
\newabbreviation{tbs}{TBS}{time-based sampling}
\newabbreviation{tgid}{TGID}{thread group ID}
\newabbreviation{tid}{TID}{thread ID}
\newabbreviation{tif}{TIF}{Top Island First}
\newabbreviation{tlb}{TLB}{Translation Lookaside Buffer}
\newabbreviation{tldk}{TLDK}{FD.io Transport Layer Development Kit}
\newabbreviation{tsc}{TSC}{Time Stamp Counter}
\newabbreviation{ue}{UE}{User Equipment}
\newabbreviation{uio}{UIO}{Userspace I/O}
\newabbreviation{upf}{UPF}{User Plane Function}
\newabbreviation{urllc}{URLLC}{Ultra-Reliable and Low-Latency Communications}
\newabbreviation{vf}{VF}{Virtual Function}
\newabbreviation{vfio}{VFIO}{Virtual Function I/O}
\newabbreviation{vim}{VIM}{Virtual Infrastructure Manager}
\newabbreviation{vm}{VM}{Virtual machine}
\newabbreviation{vmm}{VMM}{Virtual Machine Manager}
\newabbreviation{vnf}{VNF}{Virtual Network Function}
\newabbreviation{vnfc}{VNFC}{Virtual Network Function Component}
\newabbreviation{vnfm}{VNFM}{VNF Manager}
\newabbreviation{vpp}{VPP}{Vector Packet Processing}
\newabbreviation{vran}{vRAN}{Virtualized RAN} 
\newabbreviation{wcet}{WCET}{Worst-Case Execution Time}

\newabbreviation{ann}{ANN}{Artificial Neural Network}
\newabbreviation{gpuperfapi}{GPUPerfAPI}{GPU Performance API}
\newabbreviation{igpu}{iGPU}{Integrated GPU}
\newabbreviation{mape}{MAPE}{Mean Absolute Percentage Error}

\newabbreviation[
  longplural={Operating Systems},
  shortplural={OSes}
]{os}{OS}{Operating System}

\newabbreviation[
  long={\protect\ifglsused{os}{General Purpose OS}{General Purpose \glsdisp[hyper=false]{os}{Operating Systyem}}},
  longplural={\protect\ifglsused{os}{General Purpose OSes}{General Purpose \glsdisp[hyper=false]{os}{Operating Systyems}}},
  short={GPOS},
  shortplural={GPOSes},
]{gpos}{GPOS}{General Purpose Operating System}

\newabbreviation{gpgpu}{GP-GPU}{General Purpose Computing on GPU}

\newabbreviation{fsf}{FSF}{FIRST Scheduling Framework}

\setabbreviationstyle[common]{short-nolong}

\newabbreviation[category=common]{arp}{ARP}{Address Resolution Protocol}
\newabbreviation[category=common]{dhcp}{DHCP}{Dynamic Host Configuration Protocol}
\newabbreviation[category=common]{icmp}{ICMP}{Internet Control Message Protocol}
\newabbreviation[category=common]{ip}{IP}{Internet Protocol}
\newabbreviation[category=common]{nat}{NAT}{Network Address Translation}
\newabbreviation[category=common]{posix}{POSIX}{Portable Operating System Interface for Computing Environments}
\newabbreviation[category=common]{tcp}{TCP}{Transmission Control Protocol}
\newabbreviation[category=common]{udp}{UDP}{User Datagram Protocol}
\newabbreviation[category=common]{gpu}{GPU}{Graphics Processing Unit}

\setabbreviationstyle[retif-component]{long-only-short-only}

\newabbreviation[category=retif]{retif}{ReTiF}{Real-Time Framework}

\newabbreviation[
  category=retif-component,
  type=ignored-glossary,
]{retif-daemon}{ReTiF Daemon}{\emph{ReTiF Daemon}}

\newabbreviation[
  category=retif-component,
  type=ignored-glossary,
]{retif-library}{ReTiF Library}{\emph{ReTiF Library}}

\newabbreviation{ape}{APE}{Absolute Percentage Error}

\journal{Journal of Systems Architecture}

\begin{document}
\begin{frontmatter}


\author[aff_iis]{Sergio Mazzola}
\ead{smazzola@iis.ee.ethz.ch}
\author[aff_sssa]{Gabriele Ara}
\ead{gabriele.ara@santannapisa.it}
\author[aff_iis]{Thomas Benz}
\ead{tbenz@iis.ee.ethz.ch}
\author[aff_rise]{Björn Forsberg}
\ead{bjorn.forsberg@ri.se}
\author[aff_sssa]{Tommaso Cucinotta}
\ead{tommaso.cucinotta@santannapisa.it}
\author[aff_iis,aff_bol]{Luca Benini}
\ead{lbenini@iis.ee.ethz.ch}

\affiliation[aff_iis]{organization={Integrated Systems Laboratory (IIS), ETH Zürich},
            addressline={Gloriastrasse 35},
            city={Zürich},
            postcode={8092},
            country={Switzerland}}
\affiliation[aff_sssa]{organization={Real-Time Systems Laboratory (ReTiS), Scuola Superiore Sant'Anna},
            addressline={Piazza Martiri della Libertà 33},
            city={Pisa},
            postcode={56127},
            country={Italy}}
\affiliation[aff_rise]{organization={Department of Computer Science, RISE Research Institutes of Sweden},
            addressline={Isafjordsgatan 22},
            city={Kista},
            postcode={16440},
            country={Sweden}}
\affiliation[aff_bol]{organization={Department of Electrical, Electronic, and Information Engineering (DEI), University of Bologna},
            addressline={Viale Risorgimento 2},
            city={Bologna},
            postcode={40136},
            country={Italy}}

\title{Data-Driven Power Modeling and Monitoring via Hardware Performance Counter Tracking}


\begin{abstract}
  \glsresetall{}
  Energy-centric design is paramount in the current embedded computing era: use cases require increasingly high performance at an affordable power budget, often under real-time constraints. 
  Hardware heterogeneity and parallelism help address the efficiency challenge, but greatly complicate online power consumption assessments, which are essential for dynamic hardware and software stack adaptations.
  %
  \rebuttal{We introduce a novel power modeling methodology with state-of-the-art accuracy, low overhead, and high responsiveness, whose implementation does not rely on microarchitectural details.}
  Our methodology identifies the \glspl{pmc} with the highest linear correlation to the power consumption of each hardware sub-system, for each \gls{dvfs} state.
  The individual, simple models are composed into a complete model that effectively describes the power consumption of the whole system, 
  achieving high accuracy and low overhead. Our evaluation reports an average estimation error of \SI{7.5}{\percent} for power consumption and \SI{1.3}{\percent} for energy.
  %
  We integrate these models in the Linux kernel with Runmeter, an open-source, \gls{pmc}-based monitoring framework. Runmeter manages \gls{pmc} sampling and processing, enabling the execution of our power models at runtime. With a worst-case time overhead of only \SI{0.7}{\percent}, Runmeter provides responsive and accurate power measurements directly in the kernel. This information can be employed for actuation policies in \rebuttal{workload-aware \gls{dvfs}} and power-aware, closed-loop task scheduling.
\end{abstract}

\begin{keyword}
  Power modeling \sep
  runtime power estimation \sep
  embedded systems \sep
  operating systems \sep
  Linux kernel
\end{keyword}

\end{frontmatter}

\glsresetall{}


\section{Introduction}
\label{sec:introduction}

Recent years have seen a dramatic evolution in the embedded and real-time computing landscape, with increasingly demanding requirements. Applications strive for ever-higher computing capabilities and energy efficiency, pushing toward heterogeneous and massively parallel computing platforms~\cite{Cucinotta2023TECS,Alcaide20}.
However, since the end of Dennard's scaling, several \textit{walls} have been hit, from power consumption to memory, to hardware overspecialization~\cite{Fuchs2019accelwall}.
With the limits of current silicon technology exposed, pushing for maximum energy efficiency at runtime and in a dynamic fashion is paramount to meet the requirement for high performance within sustainable power budgets~\cite{duranton2021hipeac}.

The de facto standard to boost hardware power efficiency at runtime is \gls{dpm}, integrated even in today's simplest embedded systems in the form of \gls{dvfs} and clock gating. Through \gls{dpm}, different processing elements or computing islands can be independently turned off or slowed down based on the phase of the running workload.
However, to exploit the full potential of the available power knobs, the software stack must also be able to perform intelligent adaptations based on power measurements. Providing such information at the level of the OS kernel, for instance, allows the \emph{task scheduler} to perform power-aware decisions as to the assignment of computing resources to the running processes~\cite{Mascitti21}. This is essential for applications characterized by real-time constraints running on embedded systems, due to critical misprediction penalties and thermal concerns~\cite{Mascitti21,balsini2019modeling,Ara2022SAC}.

For such full-stack, energy-aware dynamic adaptations to be effective, however, \textit{accurate}, \textit{fine-grained}, and \textit{responsive} online power measurements, with negligible overhead on normal system operation, are required to close the control loop~\cite{bertran2012systematic,chau2017energy}.
An intuitive solution relies on analog power sensors. However, as discussed in \Cref{subsec:analog_power}, they unfortunately pose significant challenges in many practical scenarios~\cite{lin2020taxonomy}.
As an alternative to direct power sensing, analytical and data-driven power models have been extensively researched to obtain power measurements better suited for dynamic, online adaptations of hardware and software.
It is well-known that the \gls{pmc} activity effectively correlates to power consumption~\cite{bellosa2000benefits}, enabling accurate, data-driven power modeling for responsive and fine-grained power gauging~\cite{bertran2012systematic}. However, the complexity of selecting appropriate \glspl{pmc} with an understanding of the underlying hardware architecture, coupled with the modeling challenges of \gls{dvfs}, complicates their broader applicability in heterogeneous parallel systems with limited resources and constraining time requirements.

This paper introduces a \gls{pmc}-based approach to power consumption estimation for
modern, \gls{dvfs}-enabled, heterogeneous systems, extending our previous work on the topic~\cite{Mazzola2022SAMOS}. \rebuttal{We devise a low-overhead statistical model for the power consumption of an embedded computing system composed of a host CPU and additional specialized hardware acceleration sub-systems exposing activity counters}, a common practice in today's mobile and embedded platforms~\cite{yoon2017accurate}.
We decompose the system into its smaller, more easily approachable sub-systems, and build a lightweight \gls{lut} of simple power models, independently modeling every sub-system in each \gls{dvfs} state. The modeling simplicity of the \gls{lut} seeks to match or outperform more complex approaches to power estimation, achieving an advantageous trade-off between estimation accuracy and evaluation overhead, together with fine granularity and high responsiveness.

As a second key contribution, we propose \emph{Runmeter}, an architecture-agnostic integration strategy of the model within the Linux kernel that automatizes the collection of \gls{pmc} samples and the online evaluation of the power model in a lightweight and responsive fashion.
The approach is demonstrated with a modern, heterogeneous, \gls{dvfs}-enabled target platform, the NVIDIA Jetson AGX Xavier board. Considering its CPU and GPU sub-systems, our combined, system-level power model achieves an instantaneous power \gls{mape} of \SI{7.5}{\percent} and an overall energy estimation error of \SI{1.3}{\percent}. On this platform, the online implementation in Linux exhibits a worst-case overhead of 0.7\%, enabling the deployment of aggressive closed-loop power management strategies.

The rest of this paper is organized as follows.
\Cref{sec:relwork} frames our contributions into the context of its related work, providing background knowledge and justifications for the approach described in this paper.
\Cref{sec:model} describes our statistical power modeling approach as a generalization of \cite{Mazzola2022SAMOS}, while \Cref{sec:approach} illustrates the architecture of our novel power monitoring framework integrated within the Linux kernel.
Finally, \Cref{sec:experiments} describes the evaluation of our power model through offline validation and online evaluation.


\section{Background \& Related Work}
\label{sec:relwork}

\cite{ahmad2017survey,djedidi2020power,Zoni2023Survey} present a comprehensive survey of different power modeling approaches and runtime power monitors in the field of embedded and mobile devices.
In the following, we mainly focus on the research works related to \gls{pmc}-based power modeling and online, model-based power monitoring, comparing them with our solution. We summarize the key characteristics in \Cref{tab:soa}, providing a comparison with our approach.

\subsection{Analog Power Measurement}
\label{subsec:analog_power}

For power-aware dynamic adaptations to be possible, online power gauging is a requirement. To effectively leverage techniques such as \gls{dpm} and power-aware task scheduling, the online power gauging must possess the following properties:
\begin{enumerate}
    \item \textit{accuracy}, in terms of time resolution and sensitivity, to properly feed the power control loop;
    \item \textit{responsiveness}, to promptly reflect the hardware activity profile and provide stable feedback for the control loop;
    \item \textit{fine granularity}, in terms of introspection into the power consumption of individual hardware sub-systems (i.e., \textit{decomposability}) and task-level power budgeting.
\end{enumerate}

Several off-the-shelf \gls{soc} platforms come equipped with built-in power sensors, although not integrated on the same die of the \gls{soc} due to technological reasons. Hence, they can rarely provide the level of introspection of individual hardware sub-systems. For the same reason, off-chip parasitics pollute their measurements with longer transients, impacting the responsiveness of their measurements~\cite{Castro2018}. Their latency is further affected by the communication channel with the host, usually implemented by a serial protocol such as \glsxtrshort{i2c}, which does not match the speed of the digital domain.
Additionally, due to their physical size and deployment costs, analog gauges often suffer from limited scalability in large-scale or densely integrated systems~\cite{lin2020taxonomy}.

Although unsuitable for reliable, power-driven actuation policies, built-in analog sensors do not require external equipment for current and voltage measurements, and they can be programmatically and reliably driven. Therefore, they prove useful to build accurate, fine-grained, and responsive \gls{pmc}-based power models through the approach showcased in \Cref{sec:model}. This is demonstrated in \Cref{sec:results-runmeter}.

\subsection{\glsfmtshort{pmc}-Based Power Modeling}

\Gls{pmc}-based statistical power models have been a hot research topic for the last 20 years, spanning all computing domains from embedded computing devices at the edge to data centers in the cloud.
Being typically accessible via memory-mapped registers, \glspl{pmc} are cheap to use, and their readings are fast and reliable.
As part of the digital domain, \gls{pmc} activity promptly reflects the current state of the hardware resources, exposing desirable responsiveness properties.
\Glspl{pmc} also provide a high degree of introspection into individual hardware sub-systems~\cite{isci2003runtime}, resulting in highly decomposable \gls{pmc}-based power models~\cite{bertran2012systematic}.

However, power estimation through \glspl{pmc} raises several challenges.
Modern computer architectures, even in the embedded domain, expose hundreds of countable performance events~\cite{nvidiaxavier,armcortexA57}. Hence, the parameter selection for a robust statistical power model often requires considerable knowledge of the underlying hardware.
Growing parallelism and heterogeneity, together with the frequent lack of open documentation, further amplify the challenge.
A careful choice of the model parameters is necessary for several additional factors:
first, \glspl{pmu} can simultaneously track only a limited number or combinations of performance counters~\cite{liu2024efficient,armcortexA57,pi2019study}. Second, the amount of model predictors directly impacts evaluation overhead, which must be small for practical actuation strategies and minimal interference with the system's regular operation and time constraints. \Gls{dvfs} determines an additional layer of modeling complexity, as hardware behavior at varying frequencies has to be considered.
To the best of our knowledge, the approach we propose in \Cref{sec:model,sec:approach} is the first one to holistically address all the mentioned challenges.


\definecolor{OliveGreen}{rgb}{0.247,0.494,0.192}
\definecolor{YellowOrange}{rgb}{1,0.624,0.169}
\definecolor{OrangeRed}{rgb}{1,0.18,0.345}
\definecolor{Gray}{rgb}{0.6,0.6,0.6}

\colorlet{coloryes}{OliveGreen}
\colorlet{colorjein}{YellowOrange}
\colorlet{colorno}{OrangeRed}

\colorlet{colorverygood}{OliveGreen}
\colorlet{colorgood}{OliveGreen}
\colorlet{colorneutral}{Gray}
\colorlet{colorbad}{OrangeRed}
\colorlet{colorverybad}{OrangeRed}

\newcommand{\verygood}[1]{\textcolor{colorverygood}{#1}}
\newcommand{\good}[1]{\textcolor{colorgood}{#1}}
\newcommand{\neutral}[1]{\textcolor{colorneutral}{#1}}
\newcommand{\bad}[1]{\textcolor{colorbad}{#1}}
\newcommand{\verybad}[1]{\textcolor{colorverybad}{#1}}

\newcommand{\tworows}[2]{\begin{tabular}[c]{@{}c@{}}#1\\ #2\end{tabular}}
\newcommand{\tworowsl}[2]{\begin{tabular}[c]{@{}l@{}}#1\\ #2\end{tabular}}
\newcommand{\relwork}[3]{\vspace{1mm}\tworowsl{#1 \etal}{(#2) #3}}

\newcommand\circledsym[2]{%
  \adjustbox{height=1.15em,margin*=0 -1.25 -2.5 0}{%
    \tikz\node[circle,color=white,fill=#1,inner sep=.2pt,font=\bfseries]{#2};%
  }
}

\newcommand{\yes}{\circledsym{coloryes}{$\pmb\checkmark$}}
\newcommand{\no}{\circledsym{colorno}{$\pmb\times$}}
\newcommand{\jein}{\circledsym{colorjein}{$\pmb\approx$}}
\newcommand{\unknown}{\circledsym{colorneutral}{?}}

\newcolumntype{Y}{>{\centering\arraybackslash}X}
\newcolumntype{Z}{>{\raggedleft\arraybackslash}X}
\newcolumntype{R}{>{\adjustbox{right=7.4em,angle=310,lap=-\width+1.5em}\bgroup}c<{\egroup}}
\newcolumntype{M}{>{\adjustbox{right=7.4em,angle=310,lap=-\width+2.5em}\bgroup}c<{\egroup}}
\newcommand*\rot{\multicolumn{1}{R}}%
\newcommand*\rotm{\multicolumn{1}{M}}%
\newcommand{\tblrottitle}[1]{\rot{{#1}}}
\newcommand{\tblrottitlem}[1]{\rotm{{\makecell[cr]{#1}}}}

\newcommand{\underlinecenter}[2]{%
\setul{3pt}{.4pt}
\ul{\mbox{\hspace{#1}}#2\mbox{\hspace{#1}}}}

\begin{table}[th!]
  \caption{Comparison between representative works in the literature of \glsfmtshort{pmc}-based power modeling. \rebuttaltwo{Note that accuracy metrics, such as the power \gls{mape}, are platform- and application-dependent. Due to differences in the experimental set-ups, we report the \gls{mape} to indicate whether each approach delivers the required accuracy for the target application, rather than as a comparison across different methodologies.}}
  \label{tab:soa}
  \vspace{-0.4cm} 
  \setlength{\tabcolsep}{0pt}%
  \sisetup{range-phrase=--}%
  \center%
  \begin{tabularx}{\columnwidth}{@{}lYYYYYYYYY@{}}
    \toprule
        \parbox[t][5em][b]{1em}{}
      & \tblrottitle{Heterogeneity}
      & \tblrottitlem{Generality}
      & \tblrottitlem{Automation}
      & \tblrottitlem{\rebuttal{Min architectural}\\\rebuttal{knowledge}}
      & \tblrottitlem{Lightweight\\model}
      & \tblrottitlem{DVFS support}
      & \tblrottitlem{Decomposability}
      & \tblrottitlem{Runtime\\monitoring}
      & \tblrottitle{Power MAPE}
    \\ %
    \midrule
      \relwork{Walker}{2016}{\cite{walker2016accurate}}
      & \tworows{CPU}{only} 
      & \tworows{ARM}{cores} 
      & \yes{} 
      & \no{} 
      & \yes{} 
      & \yes{} 
      & \no{} 
      & \no{} 
      &  3-4\% 
    \\ %
      \relwork{Yoon}{2017}{\cite{yoon2017accurate}}
      & \yes{} 
      & Android 
      & \jein{} 
      & \jein{} 
      & \yes{} 
      & \yes{} 
      & \yes{} 
      & \yes{} 
      &  5.1\% 
    \\ %
      \relwork{Wang}{2019}{\cite{wang2019statistic}}
      & \tworows{iGPU}{only} 
      & \no{} 
      & \no{} 
      & \no{} 
      & \yes{} 
      & \no{} 
      & \no{} 
      & \no{} 
      & 3\% 
    \\ %
      \vspace{1.5mm}\tworowsl{Mammeri et}{al. (2019) \cite{mammeri2019performance}}
      & \yes{} 
      & mobile 
      & \jein{} 
      & \jein{} 
      & \no{} 
      & \no{} 
      & \no{} 
      & \no{} 
      & 4.5\% 
    \\ %
      \vspace{1.5mm}\tworowsl{Tarafdar et}{al. (2023) \cite{tarafdar2023power}}
      & \no{} 
      & \tworows{data}{centers} 
      & \unknown{} 
      & \no{} 
      & \yes{} 
      & \unknown{} 
      & \no{} 
      & \yes{} 
      & 4.7\% 
    \\ %
        \textbf{This work}
      & \yes{} 
      & \yes{} 
      & \yes{} 
      & \yes{} 
      & \yes{} 
      & \yes{} 
      & \yes{} 
      & \yes{} 
      & \begin{tabular}[c]{@{}c@{}}{\scriptsize CPU}\\ {\scriptsize 3-4.4\%}\\ {\scriptsize GPU}\\ {\scriptsize 6-8\%}\end{tabular} 
    \\ %
    \bottomrule
  \end{tabularx}
\end{table}

\subsubsection{\rebuttal{Bottom-up and top-down modeling}}
Bertran~\etal~\cite{bertran2013counter} identify two families of \gls{pmc}-based power models, depending on their construction: \emph{bottom-up} and \emph{top-down}.
\emph{Bottom-up} approaches rely on extensive knowledge of the underlying architecture to estimate the power consumption of individual hardware sub-systems.
Although the pioneering works of this field fall into this category~\cite{isci2003runtime,bertran2012systematic}, their results highlight the limited applicability of bottom-up power models, which are closely tied to a reference architecture. Recent research further confirms this limitation~\cite{Phung2020}.

\emph{Top-down} approaches target simple, low-overhead, and more generally applicable models, often black-boxing the platform internals. 
Over the years, this approach has been refined from the usage of few, manually selected \glspl{pmc}~\cite{bellosa2000benefits} to the employment of more elaborate procedures for \gls{pmc} selection and support for parallel~\cite{pusukuri2009methodology,singh2009real} and heterogeneous~\cite{bircher2011complete} platforms.
However, no past research investigates a combination of accurate and low-overhead models addressing \gls{dvfs} without requiring expert architectural knowledge.

In the context of CPU power modeling specifically targeting mobile and embedded platforms, Walker~\etal~\cite{walker2016accurate} employ a systematic technique for \gls{pmc} selection and train power models for the ARM A7 and A15 embedded processors. However, only one trained weight is used to estimate the power consumption at any \gls{dvfs} state. As no information on the employed \gls{dvfs} states is available, it is not possible to assess whether such a modeling choice can avoid large inaccuracies due to the limited number of parameters.

Yoon~\etal~\cite{yoon2017accurate} propose a power model for mobile \glspl{soc} solely based on the utilization metrics provided by the Android kernel, which abstracts the model deployment from the specific architecture.
However, to construct the model, the authors individually characterize each sub-system by leveraging in-depth architectural knowledge. Moreover, the single utilization parameter often fails to grasp the different phases of the running workload, \rebuttaltwo{which results in a larger estimation error with workloads showing higher variability.}
Yoon~\etal also implement an online monitor deploying their power model, which reports an overhead of up to \SI{4.5}{\%} of CPU time at a \SI{1}{\hertz} sampling frequency. In comparison, Runmeter achieves up to \SI{0.7}{\%} overhead with a sampling period of \SI{10}{\hertz}.

Top-down power modeling approaches are also applied to \glspl{gpu}.
Wang~\etal~\cite{wang2019statistic} analyze the power consumption of an AMD \gls{igpu}, carefully studying its architecture and selecting the best \glspl{pmc} to build a linear power model. While they achieve a \gls{mape} below \SI{3}{\%}, the model is manually fine-tuned for a single system, requiring its expert architectural knowledge, and \gls{dvfs} is not taken into account.
Recent works also resort to deep learning for creating accurate black-box power models: Mammeri~\etal~\cite{mammeri2019performance} train an \gls{ann} with several manually chosen CPU and \gls{gpu} \glspl{pmc}, achieving a \gls{mape} of \SI{4.5}{\%}. However, neural networks generally require a number of multiply-accumulate operations two orders of magnitude higher than for a linear model, representing a non-negligible runtime overhead. Potentially long training time, risk of overfitting, deployment challenges, and lack of decomposability are additional drawbacks of this approach.

Tarafdar~\etal~\cite{tarafdar2023power} also propose several power modeling techniques based on multi-variable linear regression, \gls{svr}, and \gls{ann}. While their approach is statistically sound, they conceive it as a solution for data centers. Therefore, no fine-grained power information about the computing platform is made available. Moreover, the model parameter selection happens a priori and is not correlated to the modeled platform. Their best power model, which features an evaluation overhead in the order of the microseconds, shows an average power estimation error of at most 4.7\%.



\subsubsection{\rebuttal{Contributions}}
Our proposed model shares its decomposability and responsiveness with bottom-up approaches but resorts to top-down modeling for individual sub-systems: \rebuttal{we trade a lower per-component introspection for a systematic modeling procedure requiring very little architectural knowledge and minimal human intervention.}
%
In addition, we carefully address the platform heterogeneity and \gls{dvfs} capabilities by introducing a \glsxtrshort{lut}-based approach that employs individual, low-overhead linear models for each sub-system and for each \gls{dvfs} state.
\rebuttal{With the aim to enable real-time, in-kernel power estimation on embedded and edge systems, simplicity, determinism, and overhead are critical. While techniques such as quantized deep learning could address inference overhead, they open up challenges in training, deployment, and reproducibility in constrained OS environments.}
On the other hand, the simplicity of linear models makes them particularly suitable for deployment in embedded systems, possibly with real-time constraints, and as part of an \gls{os} kernel, thanks to their low overhead when evaluated at runtime.

\subsection{Online Model-based Power Monitoring}

Various tools have been proposed for online, model-based estimation of power consumption through \gls{pmc} sampling.
Commonly, runtime monitoring is implemented by simply sampling the \glspl{pmc} with a fixed periodicity~\cite{Pricopi2013,Rodrigues2013}.

One of the most popular open-source tools for online \gls{pmc} sampling is \mbox{PMCTrack}, developed by Saez~\etal~\cite{saez2017pmctrack}. \mbox{PMCTrack} can monitor per-system, per-CPU, and per-process \glspl{pmc} directly within the Linux kernel.
However, targeting general-purpose use cases, many aspects of its implementation are not suited for real-time tasks, that pose different requirements from those of \ccode{SCHED\_OTHER} tasks.
As an example, \mbox{PMCTrack} may delay the generation of a \gls{pmc} sample related to a task until it is scheduled for execution by the task scheduler. Such a delay is detrimental to the responsiveness of a power-monitoring tool. \rebuttal{Other tools based on \mbox{PMCTrack} modify its source to generate samples at every context switch, albeit their target use-case again differs from that of real-time tasks~\cite{xu2021lush}, making them unsuitable for our study.}

\newcommand\runmeter{Runmeter\xspace}

On the other hand, with \runmeter, we provide a responsive and reliable mechanism to monitor the evolution of \glspl{pmc} in use cases that include real-time tasks.
As discussed in \cref{sec:approach}, we achieve this by implementing a moving sampling window for \gls{pmc} collection, featuring fully configurable time resolution and sensitivity. This results in responsive power readings that do not sacrifice estimation accuracy, which depends on the configurable window size~\cite{Rodrigues2013}.
By deploying our low-overhead power models in \runmeter, we allow the collection of online accurate power measurements with minimal overhead and sub-system-level introspection at the granularity of individual tasks.


\section{Data-Driven Power Modeling}
\label{sec:model}

\glsreset{lut}

This section describes our automated, data-driven approach to the training of a \gls{dvfs}-aware, low-overhead power estimation model for heterogeneous, embedded platforms.
\rebuttal{Given a generic target platform composed of one or multiple
sub-systems, we provide a systematic approach to its characterization (i.e., model parameter selection) based on extensive profiling of the exposed \glspl{pmc}, rather than on microarchitectural details}. This results in an accurate, responsive, and low-overhead power model for the entire platform and its individual sub-systems, at the desired operating frequencies.

\begin{figure*}[tp]
    \centering
    \hspace*{-0.16\textwidth} 
    \includegraphics{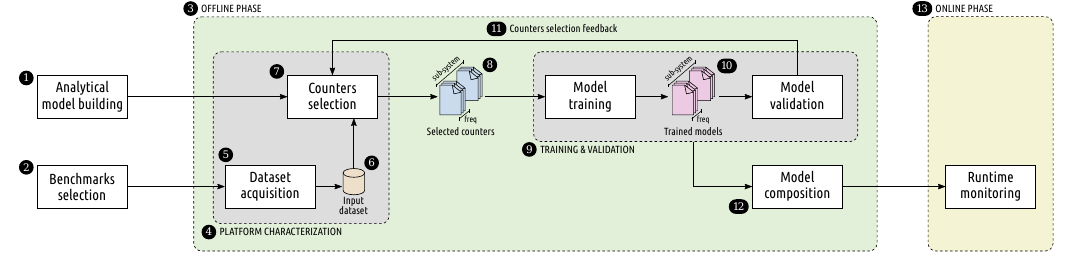}
    \caption{Scheme of the proposed data-driven, automatic power modeling approach for \glsfmtshort{dvfs}-enabled heterogeneous platforms.}
    \label{fig:pipeline}
\end{figure*}

\newcommand\circlefigref[1]{\Cref{fig:pipeline},\circleref{#1}\xspace}
\newcommand\circlefigreftwo[2]{\Cref{fig:pipeline}, \circleref{#1} and \circleref{#2}\xspace}


\subsection{The systematic, data-driven methodology}


For the purpose of this section, we consider a generic computing platform composed of a set $D$ of individual sub-systems $d$. Each sub-system $d$ has a set $F_d$ of possible \gls{dvfs} states, each one characterized by an operating frequency $f_d$.
We define $D^\ast \subseteq D$ as the subset of sub-systems that we target for power modeling. Furthermore, for each $d \in D^\ast$, we define $F_d^\ast \subseteq F_d$ as the subset of $d$'s \gls{dvfs} states that we consider. Both $D^\ast$ and $F_d^\ast$ are user-defined parameters that might vary based on the use-case.
For each sub-system $d$, up to $N_d$ distinct performance events can be tracked at the same time, i.e., $d$ features $N_d$ \glspl{pmc}.
In the following, we refer to an individual performance event exposed by the sub-system $d$ and tracked by its $i$-th \gls{pmc} as $x_{i}$, with $i = 1\,..\,N_d$. The set $X_{d, f_d}$ of all $x_{i}$ of a sub-system $d$, operating at $f_d$, represents the set of input independent variables, or the \emph{predictors}, of our models.

We propose a methodology \rebuttal{to heuristically select the best $X_{d, f_d}$ set for each sub-system/frequency in terms of overhead and accuracy}, subsequently training its related weights $W_{d, f_{d}}$. The individually generated power models $P_{d}(X_{d, f_{d}}, W_{d, f_{d}})$ are simple, linear models that we compose into a \gls{lut}, effectively grasping the different platform behaviors at varying operating frequencies.
\begin{equation}\label{eq:lut}
    LUT[d,f_d] =
    P_{d}(X_{d, f_{d}}, W_{d, f_{d}})
    \quad \textrm{for} \  d \in D^\ast, \; f_d \in F_d^\ast
\end{equation}

\noindent
The overall system power consumption $P_{tot}$ is computed by \textit{reduction sum} of the \gls{lut} along the sub-system dimension $d$, after fixing a $f_d$ for each sub-system.

\begin{equation}
    \label{eq:sysmodel}
    P_{tot} = \sum_{d\in D^\ast} LUT[d,f_d] = \sum_{d\in D^\ast} P_{d}(X_{d, f_{d}}, W_{d, f_{d}})
\end{equation}

The power consumption of a digital system has a non-linear dependency on its operating voltage and frequency. Our \gls{lut}-based approach allows us to break it down into the individual contributions of each sub-system, linearizing their power models. This greatly simplifies the model generation and evaluation, which favors its deployment within embedded, real-time systems and its applicability to different platforms~\cite{yoon2017accurate}. However, our decomposition is based on the assumption of independence among power consumption of different sub-systems. Such an assumption is justified by the limited accuracy increase of the non-linear model at the cost of a much higher overhead~\cite{mccullough2011evaluating}.

\subsection{Analytical Model Building \& Benchmarks Selection}
\label{subsec:model-benchmarks}

As a first step toward the construction of a system-level \gls{lut}, we associate the generic sub-systems $d$ with the expression $P_{d}$ of a power model based on generic performance event information (\circlefigref{1}).
Thanks to the \gls{lut}-based approach, the frequency $f_d$ is factored out. Therefore, the individual power models are reduced to linear combinations of the \gls{pmc} samples and the trained weights.
\begin{gather}\label{eq:model}
    P_{d}(X_{d, f_{d}}, W_{d, f_{d}}) = L_d + \sum_{i=1}^{N_i} \Big(\frac{1}{T}\cdot x_{i}\Big)\cdot w_{i} \\
    \textrm{for} \  d \in D^\ast, \; f_d \in F_d^\ast \ \textrm{and} \ \; \substack{L_d \\ w_{i}} \in W_{d, f_{d}}, \; x_{i} \in X_{d, f_{d}} \nonumber
\end{gather}

The weight $L_d$ is used to capture the constant component of the power consumption, i.e., the leakage, while the \gls{pmc}-dependent terms vary based on the hardware
activity, modeling the dynamic power.
The factor $\nicefrac{1}{T}$ normalizes the raw \gls{pmc} samples $x_{i}$ with respect to the sampling period $T$ of the dataset traces: training the model on \gls{pmc} rates rather than absolute values addresses sampling jitter and simplifies time-rescaling for model evaluation at arbitrary time resolutions.

The model generation is also based on the careful choice of a representative set of workloads for the platform (\circlefigref{2}). In the first place, complete coverage of all targeted sub-systems is required to address the platform heterogeneity fully.
Secondly, for each sub-system, the workloads should be diverse enough to induce a broad range of behaviors for a dataset-independent result.
The selected benchmarks are employed to build a dataset for platform characterization, model training, and model validation, containing the activity traces of the workloads, i.e., the \gls{pmc} samples.

\subsection{Platform Characterization}
\label{sec:model-characterization-steps}

Given the $P_{d}$ mathematical model for each sub-system $d$, we define \emph{platform characterization} the process of model parameter selection (\circlefigref{4}). In other words, with the platform characterization, we define the set $X_{d,f_d}$ of performance events, which will be used to model the dynamic power of each sub-system $d$ at each frequency $f_d$.

Individually for each sub-system $d$ and frequency $f_d$, we perform a one-time correlation analysis between all of its local \glspl{pmc} and the sub-system power consumption, \rebuttal{looking for the $X_{d, f_{d}}$ that achieves the most convenient trade-off in terms of model accuracy and estimation overhead under the constraints imposed by the \gls{pmu} limitations.} Note that different frequencies of the same sub-system $d$ might be assigned with different \glspl{pmc}, which effectively models \gls{dvfs} with simpler, linear models.
\rebuttal{This characterization process is meant as an automatic, data-driven alternative to methodologies requiring expert microarchitectural knowledge, such as manual \gls{pmc} selection and analytical power modeling~\cite{leng2013gpuwattch}.}
Given the sub-system $d$, its characterization involves the following steps:

\begin{enumerate}
    \item for each \gls{dvfs} state $f_d \in F_d^\ast$, we profile \emph{all} performance events exposed by $d$ while tracing $d$'s power (\circlefigref{5}); \rebuttal{as time-correlated \gls{pmc} and power measures are only required for post-mortem traces, simple synchronization techniques can be used depending on the nature of the power sensor~\cite{bircher2011complete,walker2016accurate};}

    \item we normalize the \gls{pmc} samples with respect to the sampling periods to overcome sampling jitter;

    \item we compute a \gls{lls} regression of each event's activity trace over its related power measurements, for each $f_d$; we discard events with a p-value above \textit{0.05} as not reliable for a linear correlation;

    \item individually for each $f_d$, we sort the remaining events by their \gls{pcc} and select the best-scoring ones that can be profiled simultaneously (\circlefigreftwo{7}{8}).
\end{enumerate}

To compose $X_{d, f_{d}}$, it is usually enough to select the desired number of best-scoring performance events. \rebuttal{A higher number of model parameters, within the limit of overfitting, usually corresponds to higher model accuracy, but also larger evaluation latency}. The optimal number of \glspl{pmc} with respect to model accuracy can be defined by iteratively considering the estimation error results from the model evaluation step (\circlefigref{11}).
\rebuttal{
On the other hand, this step has to consider the limitations of the platform's \gls{pmu}, which come in two forms:
\begin{itemize}
    \item \glspl{pmu} have a limited number of \glspl{pmc} to track events. For example, typical embedded ARM CPUs feature up to four or six individual counters, that can be mapped to freely selectable or fixed performance events. High-performance CPUs, like the \emph{Hisilicon Kunpeng 920} processor, can track up to twelve events per core domain~\cite{liu2024efficient}.
    \item some performance events are mutually exclusive, i.e., \textit{incompatible}.
\end{itemize}
Incompatible events can be tracked in a time-sharing fashion through \emph{counter multiplexing}. However, such an approach increases tracking overhead and introduces interpolation errors, decreasing estimation accuracy~\cite{liu2024efficient}. Targeting low-overhead, real-time model estimation, we opt for a lower-complexity solution that avoids \gls{pmc} multiplexing. To this end, we devise a \gls{pmu}-aware heuristic identifying the subset of compatible counters that provide the highest power estimation accuracy while complying with \gls{pmu} constraints (\circlefigref{11})~\cite{Mazzola2022SAMOS}.}

\subsection{Training, Validation, and System-level Model}
\label{sec:model-train-validation-composition}

With the sets of counters $X_{d, f_{d}}$ defined during platform characterization, we compose the \gls{lut} of \Cref{eq:lut} by individually training the linear power model $P_{d}(X_{d, f_{d}}, W_{d, f_{d}})$ of each sub-system $d \in D^\ast$ for each $f_d \in F_d^\ast$ (\circlefigref{9}). The output of each training step is a set of weights $W_{d, f_d}$ (\circlefigref{10}).
To train each individual $P_{d}(X_{d, f_{d}}, W_{d, f_{d}})$, we perform a \gls{nnls} linear regression of the \glspl{pmc} rates
over the power measurements, obtaining the set of non-negative weights $W_{d, f_{d}}$. Compared to unconstrained \gls{lls}, non-negative weights are physically meaningful and prove to be robust to multicollinearity, which makes our simple models less prone to overfitting.
We subsequently validate each individual $P_{d}(X_{d, f_{d}}, W_{d, f_{d}})$.

After individual training and validation, we combine all the individual sub-system models (\circlefigref{10}) into the system-level power model (\circlefigref{12}) defined by \cref{eq:sysmodel}. Under the reasonable assumption of power consumption independence among sub-systems~\cite{mccullough2011evaluating}, such an approach relieves us from profiling all possible combinations of sub-systems' frequencies. This simplifies and accelerates the platform characterization, ultimately generating a simple linear model that is more robust to overfitting, accurate, lightweight, and decomposable.
%

The complete model can finally be used online (\circlefigref{13}) to monitor the instantaneous power consumption of the entire system and its sub-systems. To do so, it is enough to keep the weights $W_{d, f_{d}}$ available at runtime for each $(d,\,f_d)$ combination of interest. Due to our modeling methodology, this has a negligible memory footprint. After acquiring the \gls{pmc} samples for the set of model parameters $X_{d, f_{d}}$, the power model can be efficiently evaluated with the small number of multiply-accumulate operations required by its linear expression. Our proposed framework for the online deployment of our power model is the object of the next section.


\newcommand\runmeterframework{\runmeter Framework\xspace}
\newcommand\runmetermodule{\runmeter Kernel Module\xspace}
\newcommand\runmeterpatch{\runmeter Kernel Patch\xspace}

\section{Online Monitoring and Kernel Support}
\label{sec:approach}

The target of our work is to enable power awareness in crucial functionalities, such as task scheduling and resource allocation, when running with real-time applications on resource-constrained devices.
The modeling approach discussed in \Cref{sec:model} results in a complete system-level power model that can provide accurate and introspective power estimates with minimal overhead.
An online monitoring framework that integrates the proposed model is also essential to our target. For this, we propose \runmeter, an online monitoring framework integrated in the Linux kernel\footnote{\rebuttal{
\runmeterframework is an open source project; its homepage is at \url{https://gitlab.retis.santannapisa.it/ampere/runmeter}. Here, users can find all the necessary tools to build and deploy the framework on supported platforms.
}}.

\runmeter supports the runtime estimation of system- and task-level metrics, including power and energy consumption, through \gls{pmc} tracking. As such, it provides the infrastructure to flexibly collect \gls{pmc} samples with minimal overhead and evaluate the model presented in Section \ref{sec:model}, exposing its estimates to the Linux scheduler.
\rebuttal{The modular design of \runmeter abstracts its implementation from the specifics of the hardware architecture}. Only a minimal subset of its components must be re-implemented to support different target platforms, sub-systems, and tracked metrics.
As a case study, we leverage the framework to implement support for online power estimation and monitoring of the CPU sub-system.

\subsection{\rebuttal{\runmetermodule}}
\label{subsec:runtime-patch}
\label{subsec:runtime-module}



Once loaded into the kernel, the \runmetermodule hooks to strategic callbacks to trace and collect running statistics on a selection of the available \glspl{pmc}, according to the result of the platform characterization (\Cref{sec:model-characterization-steps}). Since a different set of counters $X_{\text{CPU},f_{\text{CPU}}}$ can be selected to model the evolution of the platform depending on the \gls{dvfs} state $f_{\text{CPU}}$, the module selects the correct \glspl{pmc} to track according to the model \gls{lut} (\cref{eq:lut}).
The module also subscribes to the CPU frequency governor (\ccode{CPUFreq}) to be notified of each change of frequency so that it can dynamically reconfigure the set of tracked \glspl{pmc} for each CPU core.
When tracking is enabled, the kernel module generates a new \gls{pmc} sample on each CPU core whenever one of the following events occurs:
\begin{itemize}[noitemsep,nolistsep]
    \item a context switch, in which case a new sample is always generated, or
    \item a user-configurable number of scheduler ticks since the last sample was produced on that core.
\end{itemize}
The first trigger allows \runmeter to collect \gls{pmc} statistic on a per-task basis and derive power estimates with task-level granularity.
The second trigger, on the other hand, provides an upper bound to the inter-arrival time between two consecutive \gls{pmc} samples. This guarantees that tasks hogging the CPU do not interfere with the monitoring. Since this bound is expressed in terms of scheduler ticks, its granularity depends on the \ccode{CONFIG\_HZ} Linux kernel option. \rebuttal{The dependence on scheduler ticks also prevents unwanted activation of Runmeter during deep idle states.}

Proper selection of this upper bound is key to ensuring the desired responsiveness when monitoring CPU counters. The basic \gls{pmc} sampling mechanism provides the accumulated value of the event counter since its last reading.
In this case, a small sampling period negatively impacts the information collected by the \glspl{pmc}: since each sample is tied to a single task, it is difficult to derive any meaningful data about the overall platform status from it.
On the other hand, with a large sampling period, the read-out value is updated less frequently, which is detrimental for actuation policies requiring high responsiveness~\cite{Rodrigues2013}.

As a trade-off, we devise a \emph{moving-window} approach that decouples the \gls{pmc} sampling period from their observation window. The moving window allows us to obtain \gls{pmc} statistics accumulated over an arbitrarily long window and updated at an arbitrarily fine time granularity.
To implement the moving-window approach, we instantiate a \emph{window buffer} for each \gls{pmc}. Each buffer stores a user-configurable number of the most recent \gls{pmc} samples. 
%
The value of each \gls{pmc} over the whole window is tracked by summing up all samples in the buffer. This information is updated each time the window moves forward (i.e., a new sample is available). We refer to this value as \emph{synthetic \gls{pmc} sample}. Such a moving buffer also serves the purpose of aggregating the per-task \gls{pmc} samples to obtain core-level metrics.

Consuming synthetic samples provides more meaningful \gls{pmc} data for the metrics to be estimated. This comes at the cost of the additional processing of each \gls{pmc}'s window.
\rebuttal{
However, as shown in \Cref{sec:experiments}, this additional processing introduces negligible overhead in practice. Moreover, the aggregation of samples at the core level is required regardless of the moving-window mechanism, making the relative cost of this step minimal.
}

\subsection{In-Kernel CPU Power Model}
\label{subsec:cpu_model}

Synthetic \gls{pmc} samples provide visibility over a time defined by the window size, but are updated at a rate defined by the \gls{pmc} sampling period.
Components like an online CPU power (or energy) monitor are required to re-evaluate their estimates at each update of the synthetic samples. A high degree of responsiveness in the monitoring, useful for robust actuation, can be achieved when the model evaluation time can keep up with the stream of synthetic samples.

The model we present in \Cref{sec:model} retains high accuracy despite its low computational complexity, which empowers it to actually keep up with the stream of samples. As a case study, we deploy it in the Linux kernel through the infrastructure provided by the \runmeterframework.
The CPU power monitor in \runmeter implements, for each CPU \gls{dvfs} state, the following model, extending \Cref{eq:model} to support multiple CPU cores:
\begin{equation}\label{eq:cpu_power_model}
\begin{aligned}
    P_{\text{CPU}} = L_{\text{CPU}} + 
    \underbrace{\sum_{i = 1}^{\#\text{cores}} \sum_{j = 1}^{N_\text{CPU}} \Big(\frac{1}{T'} \cdot x_{ij}\Big) \cdot w_{ij}}_{\displaystyle = \; \frac{1}{T'} \sum_{i = 1}^{\#\text{cores}} \sum_{j = 1}^{N_\text{CPU}} x_{ij} \cdot w_{ij}}
\end{aligned}
\end{equation}
\noindent
The weights $L_{\text{CPU}}$ and $w_{ij}$ are fractional values, but the usage of floating-point arithmetic within the Linux kernel is problematic and expensive.
For this reason, we use fixed-point arithmetic to implement the in-kernel power model, supported by the negligible loss of dynamic range and precision that we evaluate in \Cref{sec:results-runmeter}.

The factor $\nicefrac{1}{T'}$ normalizes the value of each synthetic sample with respect to the width of the user-configured observation window $T'$. $T'$ might indeed differ from the sampling period $T$ of the model training dataset (\Cref{eq:model}).
Thanks to the linearity of our models, we can perform the normalization by factoring out $\nicefrac{1}{T'}$ and operating only one multiplication after the summation. This achieves arbitrary time-rescaling of the model with negligible overhead.

\subsection{\rebuttal{\runmeter and PMCTrack}}

\rebuttal{

\runmeterframework's kernel components are based on the implementation of \mbox{PMCTrack}~\cite{saez2017pmctrack}, albeit some fundamental differences branch away from the original implementation due to the specific requirements of our use case. This section clarifies the similarities and the key differences between the two tools, motivating our design choice.

Runmeter's current implementation exploits a kernel patch to insert callbacks to its \gls{pmc} sampling mechanism, detailed in \Cref{subsec:runtime-module}. We implement the rest of the \runmeterframework as a dynamically loadable kernel module that hooks to such entry points.
The \runmeterpatch is equivalent to the one provided by \mbox{PMCTrack}. \mbox{PMCTrack} additionally provides a mechanism to dynamically inject entry points for the kernel module without a kernel patch~\cite{bilbao2023flexible}. Such a mechanism is based on \emph{dynamic ftrace}, which, from Linux kernel v5.9 on, provides a stable interface to inject the hooks required by \mbox{PMCTrack}.
As reported in~\cref{sec:experiments}, our target platform relies on the Linux kernel v4.9, which therefore requires a patch to support the \runmeter kernel module. Nevertheless, \runmeter can seamlessly leverage the very same mechanism for a patch-less implementation with more recent kernel versions.

On the other hand, \mbox{PMCTrack} and \runmeterframework substantially differ in their respective kernel modules. \mbox{PMCTrack}'s users can register \emph{monitoring modules} as consumers of \mbox{PMCTrack}-managed \gls{pmc} samples~\cite{saez2017pmctrack}. However, \mbox{PMCTrack}'s limited tracing modes prevent us from implementing \runmeter as a monitoring module for \mbox{PMCTrack}.
In particular, \mbox{PMCTrack} implements three different tracing modes: \begin{enumerate*}
    \item an \emph{event-based mode}, that generates \gls{pmc} samples when one of the counters reaches a configured threshold, and
    \item two different variants of a \emph{timer-based mode}, called \gls{tbs}, that generate samples periodically.
\end{enumerate*}
\rebuttaltwo{Due to the use case targeted by \runmeter, in this work we focus on a timer-based sampling approach.}

The first \gls{tbs} mode provided by \mbox{PMCTrack} is \textit{per-task tracing}.
Per-task tracing generates new \gls{pmc} samples periodically, based on a callback executed by the scheduler tick (as in \runmeter), or when tasks are selected for execution, rather than switched out. In the latter case, the \gls{pmc} sample generation is delayed until the task is selected again.
Furthermore, each traced task has to be configured individually, which makes it challenging to trace all the tasks running on the system. While this enables \mbox{PMCTrack} to profile different performance events for each task, it represents a drawback for \runmeter, whose event selection is dictated exclusively by the operating frequency.
%
%
This limitation is common to many tracing tools~\cite{xu2021lush}. \runmeter, on the other hand, targets a rapid swap among different sets of \glspl{pmc} whenever the operating frequency changes, to align to the power model's \gls{lut} (\Cref{eq:lut}) determined during the platform characterization without looping through all the individually traced tasks.

\mbox{PMCTrack} also implements a system-wide \gls{tbs} mode, where a kernel timer dictates the sampling of \glspl{pmc} on a per-core basis~\cite{saez2017pmctrack}. This mode provides a predictable sampling period suitable for real-time power estimation, but it generates \gls{pmc} samples on a per-core basis. This prevents monitoring modules from collecting task-level metrics. In contrast, \runmeter's hybrid sampling strategy generates \gls{pmc} samples on a per-task basis, subsequently aggregating them into per-core samples, granting a predictable periodicity in the \gls{pmc} collection.
Furthermore, being driven by scheduler ticks rather than a kernel timer, \runmeter periodic activation would not wake up a core in a deep idle state, polluting power measurements, as opposed to PMCTrack system-wide \gls{tbs} mode.


Motivated by our requirements, we leverage \mbox{PMCTrack}'s abstraction of the low-level \gls{pmc} sampling and execution hooks while re-implementing the \gls{pmc} collection and delivery to consumers, tailoring it to our specific requirements.

}


\section{Evaluation}
\label{sec:experiments}

\glsreset{mape}

In this section, we evaluate the holistic power modeling approach discussed in \Cref{sec:model}, and its in-kernel implementation within the \runmeter online monitoring framework described in \Cref{sec:approach}.

\subsection{Experimental methodology}
\label{subsec:exp-method}

The target platform for our experiments is an \xavierboard, powered by the Xavier \gls{soc}~\cite{nvidiaxavier}. It is a highly parallel and heterogeneous \gls{soc} provided with an 8-core 64-bit ARMv8.2 CPU, a 512-core NVIDIA Volta \gls{gpu}, and several additional accelerators for deep-learning, computer vision, and video encoding/decoding. With many \gls{dvfs} states available for its sub-systems, this platform represents a challenging state-of-the-art target to validate our approach.
In particular, the single CPU island on the platform can be clocked at 29 different discrete frequencies between \SI{115}{\mhz} and \SI{2.3}{\giga\hertz}, while the \gls{gpu} has 14 available \gls{dvfs} states between \SI{115}{\mhz} and \SI{1.4}{\giga\hertz}.
For the evaluation of our power modeling approach, we target the CPU and \gls{gpu} sub-systems,
$$D^\ast = \{CPU,\ GPU\}$$
considering the following \gls{dvfs} states:
\begin{gather*}
    F_{\text{CPU}}^\ast = \{ \SI{730}{\mega\hertz},\; \SI{1.2}{\giga\hertz},\; \SI{2.3}{\giga\hertz} \}
    \\
    F_{GPU}^\ast = F_{GPU} = \{ \text{all 14 from \SI{115}{\mega\hertz} to \SI{1.4}{\giga\hertz}} \}
\end{gather*}
To build the input dataset, we profile several workloads based on the considerations of \Cref{subsec:model-benchmarks}. For the CPU, we employ the OpenMP benchmarks from the Rodinia 3.1 heterogeneous benchmark suite~\cite{che2009rodinia} in several multi-thread configurations, \rebuttal{in addition to further synthetic benchmarks targeting static stress test of common compute- and memory-bound patterns, such as \texttt{memcpy}~\cite{Mazzola2022SAMOS}}. For the \gls{gpu}, we employ the CUDA benchmarks from Rodinia. To average out possible interference in our measurements, such as unpredictable \gls{os} activity, each workload is profiled 3 times.

\gls{pmc} samples are acquired in a continuous, periodical mode with a sampling period of \SI{100}{\milli\second}. During each sampling period,  power measures of the CPU and GPU sub-systems are also acquired from the INA3221 built-in power monitors~\cite{ina3221}. This grants the time correlation needed for an effective correlation analysis and training~\cite{malony2011parallel}. We find that collecting more than one sample per \SI{100}{\milli\second} does not capture any additional information due to the electrical inertia of the built-in current sensors.

As discussed in \Cref{sec:relwork}, typically, built-in power monitors are not robust tools for online, power-aware actuation policies. This is mainly due to their speed, coarse granularity, and low resolution, which for the Xavier is limited to about \SI{200}{\milli\watt}.
However, they are helpful for building datasets to achieve higher introspection, time granularity, and responsiveness enabled by \gls{pmc}-based power models, as proved in \Cref{sec:experiments}.

\subsection{Offline Platform Characterization and Modeling}
\label{sec:results-characterization}

This section discusses the result of the power model generation (\Cref{sec:model}) of the individual CPU and GPU sub-systems for the NVIDIA Jetson AGX Xavier case study.

\subsubsection{Sub-system Characterization}
\label{subsec:platform_char}

For the CPU sub-system, the results of the platform characterization suggest that the power consumption of the cores is highly correlated, depending on the selected \gls{dvfs} state, with the number of cycles during which the cores are not power-gated, the number of retired instructions, the floating point activity, and various cache-related events.
The ARM \gls{pmu} always exposes the CPU active cycle counter. For the remaining power model parameters, we consider the three best counters for each frequency, as the maximum allowed by the \gls{pmu}. From our experiments, all selected performance events are compatible with each other.

For the \gls{gpu} sub-system, our results expose multiple incompatibilities among the performance events that best correlate with the power profile. To be able to simultaneously track the best model parameters at runtime, we adopt the \gls{pmu}-aware heuristic described in \Cref{sec:model-characterization-steps}, to identify the viable set of compatible performance events.
We conclude that events related to L2 cache utilization and warp execution best correlate with the power consumption of the \gls{gpu}. Additionally, through the validation step, we find that a number of eight \glspl{pmc} per frequency is the optimal trade-off between model evaluation time and the power estimate accuracy.

\subsubsection{Sub-system Modeling and Validation}
\label{subsec:train_valid}

\begin{figure*}[th!]
    \centering
    \includegraphics[trim={0 0.25cm 9.4cm 0},clip]{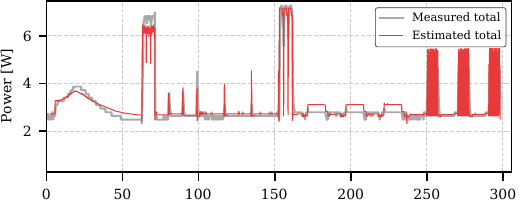}
    \includegraphics[trim={9.4cm 0 0 0},clip]{model_eval.pdf}
    \caption{Instantaneous power estimate over the validation set for the system-level power model (on the left) and its breakdown into the individual sub-system estimates (on the right), with $f_{\text{CPU}}=\SI{1.2}{\giga\hertz}$, $f_{\text{GPU}}=\SI{830}{\mhz}$.}
    \label{fig:comb_model_power}
\end{figure*}

For the CPU, we adopt a \gls{nnls} regression to individually train a linear model based on \Cref{eq:lut} for each frequency. We employ four independent variables per core, i.e., the three configurable \glspl{pmc} for each frequency and the cycle counter. Out of our input dataset, we use a random selection of \SI{70}{\%} of the total data for training and the remaining \SI{30}{\%} for validation. In terms of instantaneous power accuracy, the model achieves a \gls{mape} between \SI{3}{\%} and \SI{4.4}{\%} based on the frequency, with a standard deviation of approximately \SI{5}{\%}. When employed to estimate the energy over the entire validation set, our model achieves a maximum error of \SI{4}{\%}, \rebuttal{delivering an equal or higher accuracy as reported by the previous state-of-the-art work. It shall be noted that direct comparisons on the same hardware would require adapting and updating the previous works to the platform targeted by our experiments. These adaptations could determine accuracy degradations, whose attribution to limitations of the original methodologies could be debatable. On the other hand, \rebuttaltwo{as stated in \cref{tab:soa},} comparing the results achieved on the original targets serves as an indication that the proposed automatic and data-driven methodology delivers results within the expected level of accuracy for relevant applications, e.g., power-aware task scheduling.}

For the \gls{gpu}, we likewise train the \Cref{eq:lut} for each of the 14 \gls{gpu} frequencies with a \gls{nnls} linear regression. We use a 70\% and 30\% ratio for the training and validation set. Comparing the instantaneous power consumption estimation with the data measured on the real platform, we obtain a \gls{mape} between \SI{6}{\%} and \SI{8}{\%}, depending on the frequency. The standard deviation over all frequencies is approximately \SI{8}{\%}. The maximum energy estimation error over the full validation set is \SI{5.5}{\%} over all frequencies, with an average of \SI{2.2}{\%}.

\subsection{Combined Model Evaluation}
\label{subsec:comb_model_res}

After building, training, and validating the CPU and \gls{gpu} power models individually, we combine them to obtain a system-level power model for every possible combination of $f_{\text{CPU}} \in F_{\text{CPU}}^\ast$ and $f_{\text{GPU}} \in F_{\text{GPU}}^\ast$, corresponding to the LUT of \Cref{eq:lut}.
%
\Cref{fig:comb_model_power} shows how our decomposable power model can effectively track the instantaneous power consumption of the system over time. The achieved instantaneous power \gls{mape} of the final, combined model has an average of \SI{8.6}{\%} over all CPU and \gls{gpu} frequency combinations.
Regarding energy, the model reaches an average estimation error of \SI{2.5}{\%}.

Our results highlight that the estimation error of the combined model is higher when $f_{CPU}$ and $f_{GPU}$ diverge from each other. In particular, when $f_{GPU}$ is very low compared to $f_{CPU}$, the CPU may stall waiting for the offloaded computation. Our power model is not capable of capturing such behavior, which depends on the interaction among different sub-systems, due to our assumption of sub-system independence (cf. \cref{sec:model}). On the other hand, a real use case where such a scenario occurs is highly unlikely due to its inefficiency.
As a consequence, restricting our evaluation to real scenarios, our assumption of sub-system independence is still valid: by considering only reasonably close CPU and GPU frequencies, in particular $f_{\text{GPU}}>\SI{600}{\mhz}$, we report an instantaneous power \gls{mape} of \SI{7.5}{\%} and energy estimation error of \SI{1.3}{\%}, with a maximum of \SI{3.1}{\%}.

\rebuttal{
This accuracy must be interpreted relative to the precision of our INA3221 reference sensor, which itself has a power resolution of \SI{200}{\milli\watt} (around \SI{3}{\percent} of the measurements in our experiments)~\cite{ina3221}.
Beyond these quantitative results, it is important to acknowledge further practical sources of uncertainty that inherently limit the achievable accuracy of \gls{pmc}‑based models, along with possible mitigations.
Because we train our models directly on the target device, fabrication-time process variations and normal supply‑voltage fluctuations are inherently captured. Likewise, temperature‑dependent leakage is reflected in our dataset, and we observe negligible drift across the temperature swings imposed during the intensive platform characterization and model training. Finally, longer‑term aging effects, which develop over months or years, can be corrected via periodic retraining using our lightweight, automatic approach without requiring a full recharacterization.
}


\subsection{CPU Power Monitoring with Runmeter}
\label{sec:results-runmeter}

To evaluate our online monitoring framework, \runmeter, we integrate it into the kernel of the Linux distribution running on the NVIDIA Jetson AGX Xavier. Then, through \runmeter, we implement the power monitor discussed in \Cref{sec:approach} with support for the CPU sub-system, collecting \gls{pmc} samples with \SI{10}{\hertz} sampling period.
\rebuttal{
In all experiments, the target platform uses a patched version of the NVIDIA Jetson Linux kernel that includes the entry points for the \runmeter module. The most recent version of said kernel at the time of our evaluation is v4.9.253\footnote{\url{https://gitlab.retis.santannapisa.it/ampere/runmeter/kernel-jetson}}.
}

In this section, we discuss the impact of the fixed-point implementation of our power model, which is necessary for the in-kernel implementation. Subsequently, we validate the power estimations resulting from the online monitoring and evaluate its overhead.

\subsubsection{Fixed-point Approximation Error}
\label{subsec:exp-fixp}

In this section, we evaluate the approximation error introduced by our fixed-point implementation of the power model described in \Cref{sec:approach}, necessary to integrate it as part of a Linux kernel module. For the fixed-point implementation, we use 64-bit integers, assigning the $29$ less significant bits to the fractional part.
To analyze the approximation error, we collect the data published by the \runmetermodule and feed them to 
a user-space C++ checker procedure. The checker evaluates the model through the floating-point and the fixed-point implementations, measuring the deviation. For this evaluation, we use the same validation set discussed in \Cref{sec:results-characterization}.

\begin{figure}[t]
    \centering
    \includegraphics[width=0.8\textwidth]{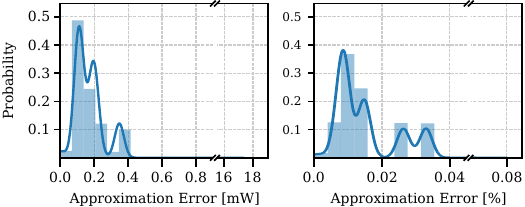}

    \caption{
        Distribution of approximation error between floating-point and fixed-point implementations of the CPU power model. The distribution is shown in terms of both absolute power approximation error and percentage error.
    }
    \label{fig:fixp-error}

\end{figure}

\Cref{fig:fixp-error} shows the distribution of such a deviation.
From our extensive evaluation, the maximum absolute approximation error is about \SI{17}{\milli\watt}; the mean error, however, is only of about \SI{0.17}{\milli\watt}. The maximum percentage error is always below \SI{0.8}{\%} of the power consumption estimated using floating-point arithmetic, with a mean error of about \SI{0.015}{\%}.
Given the negligible magnitude of the error introduced by the fixed-point implementation, we conclude that this approximation does not impact the accuracy of the model in any meaningful way.

\subsubsection{Online Power Estimation Accuracy}
\label{subsec:exp-runmeter}

Employing the fixed-point implementation of the CPU power model validated in the previous section, we integrate the power monitor in the \runmeterframework.
We then log the online estimates computed at runtime by the power model to later perform post-mortem analysis.
Therefore, differently from what analyzed so far, the power estimates reported in this section are computed directly at runtime as soon as new \gls{pmc} samples are available. The profiled workloads are the same as the validation set discussed in \Cref{sec:results-characterization}.

\begin{figure}[tp]
    \centering
    \includegraphics{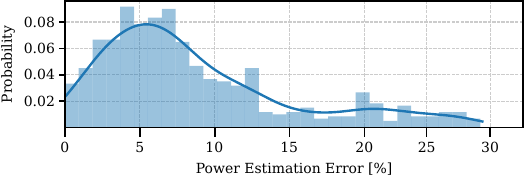}
    \caption{Distribution of the \gls{ape} of the online energy estimates over the duration of each benchmark.}
    \label{fig:energy-error-distribution}
\end{figure}

\Cref{fig:energy-error-distribution} shows the \gls{ape} distribution of the energy estimation provided by the in-kernel model when compared against the value collected from the onboard analog sensor.
The maximum \gls{ape} registered over all our experiments is around \SI{29}{\%}, the error at \nth{90} percentile is around \SI{20.8}{\%}, and the \gls{mape} is around \SI{9}{\%}.

\begin{figure}[t]
    \centering
    \subcaptionbox{\SI{730}{\mega\hertz}\label{fig:cpu_online_730}}{
        \includegraphics[]{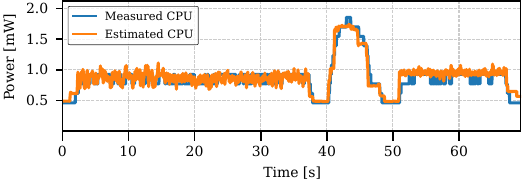}
    }

    \vspace*{.5em}

    \subcaptionbox{\SI{1.2}{\giga\hertz}\label{fig:cpu_online_1200}}{
        \includegraphics[]{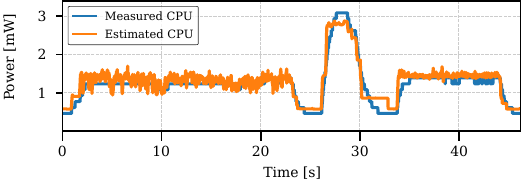}
    }

    \vspace*{.5em}

    \subcaptionbox{\SI{2.3}{\giga\hertz}\label{fig:cpu_online_2300}}{
        \includegraphics[]{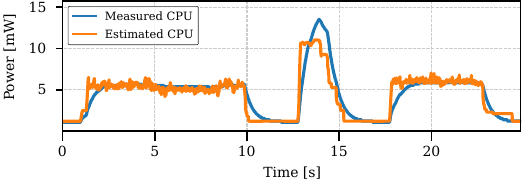}
    }

    \caption{
        Comparison of the instantaneous CPU power consumption measurement provided by the onboard INA3221 sensor and the estimation computed at runtime by the in-kernel power model. Each plot represents the same sequential execution of several workloads over time at different frequencies.
    }
    \label{fig:energy-estimation}
\end{figure}

According to our experiments, the majority of the estimation error accounted for during such an evaluation is to be attributed to very specific time frames when the phase of the workload abruptly changes.
Such behavior is visible in the example CPU power profiles depicted in \Cref{fig:energy-estimation}.
On sharp changes in the system activity, inducing rapid switches in the power consumption, the power estimated by the \gls{pmc}-based power model has faster rising and falling edges than the power measured by the analog sensor. This is especially visible at higher CPU frequencies, where the inertia of the analog current sensors has an increasingly worse impact on the latency of the measurements.
On the other hand, \glspl{pmc} are embedded in the digital domain, and their values instantly reflect the dynamic behavior of the monitored workloads.
Nevertheless, our power modeling approach makes use of the onboard power sensor to build the input dataset for training and validation. While this makes the procedure automatic, easier, and less error-prone, it creates an unavoidable discrepancy between our estimates and the ground truth during transients, related to the high responsiveness of the \gls{pmc}-based model.

The problem emerges, then, of how to assess the reliability of our power estimates during transients if the model has been trained with a built-in analog sensor. As a solution, during the training phase, we deliberately bias the training set toward workloads with more stable activity: this means that the power model is trained, on average, with power values matching the actual consumption of the platform. Thanks to the linearity of the power models, such a solution decouples the trained weights from the low sensibility and time granularity of the input power data used for training. Once trained, the power model can scale and interpolate those values according to the \gls{pmc} samples collected at runtime, providing faster power estimation and higher responsiveness.

\subsubsection{Monitoring Overhead}
\label{subsec:exp-overhead}

\runmeter is integrated with callbacks triggered at specific times during the Linux kernel execution. This imposes a certain processing overhead mainly due to \gls{pmc} data collection and manipulation, including model estimation.
To measure it, we profile the execution of the \runmetermodule callbacks. We perform these measurements in various working conditions, ranging from an \quotes{idle} state to the execution of multiple parallel applications from the set of benchmarks described in
\cite{Mazzola2022SAMOS}. We use the same frequencies employed for the CPU model evaluation.

The maximum overhead is reported when many applications execute concurrently on the system, as the number of invocations of \runmeter's callbacks increases with the number of context switches performed by the system.
In the worst-case condition of intense context switching, the time spent executing all of the framework's callbacks never exceeds \SI{7}{\milli\second} per second (i.e., \SI{0.7}{\%} overhead). Moreover, the execution of all framework's callbacks significantly speeds up when increasing the CPU frequency, reducing to less than \SI{2}{\milli\second} per second in the worst case (\SI{0.2}{\%} overhead) when operating at \SI{2.3}{\giga\hertz}.
In idle conditions, the overhead of the framework at \SI{2.3}{\giga\hertz} is always less than \SI{0.4}{\milli\second} per second (\SI{0.04}{\%}).


\section{Conclusions and Future Work}
\label{sec:conclusions}

\rebuttal{With this work, we propose a systematic, data-driven approach to \gls{dvfs}-aware statistical power modeling of heterogeneous computing systems, whose
implementation is decoupled from the target platform's microarchitectural details}. We individually model each sub-system through its local \glspl{pmc}, autonomously selecting the best ones to represent its power consumption. The sub-system models are later composed in a LUT-based system-level power model, able to grasp the complex behaviors of \gls{dvfs}-enabled hardware using simple, linear expressions. \rebuttal{This approach achieves a novel combination of automated model construction, low-overhead evaluation, high accuracy, responsiveness, and decomposability, proving itself suitable for real-time applications running on mobile and embedded systems.}

To demonstrate the applicability of our power model, we propose \runmeter, a flexible framework for \gls{pmc} monitoring and power model evaluation from within the Linux kernel. \runmeter is a substantial improvement over existing mechanisms based on \gls{pmc} tracking, as it focuses on minimizing the response time between \gls{pmc} observation and model evaluation, enhancing the responsiveness of power estimates with negligible overhead.

The validation of our power modeling approach on the state-of-the-art \xavierboard embedded platform results in power and energy estimation accuracies aligned with or \rebuttal{higher than reported by previous state-of-the-art work}.
By integrating \runmeter in the Linux kernel of the same platform, we also prove the viability of our modeling and monitoring approach for online power tracking, a key prerequisite to implementing robust power-aware control loops in \gls{dpm} and power-aware task scheduling.

\rebuttal{While our methodology is designed to be independent of microarchitectural details, future work will explore its validation across multiple hardware platforms to further reinforce its general applicability. \rebuttaltwo{The automatically identified \gls{pmc} set can also serve the development of an initial model, which can subsequently be refined with expert architectural knowledge when tighter accuracy constraints are required for a given application.}
Exploring different target platforms will also serve to demonstrate additional capabilities of our power models, such as leveraging \emph{deep idle states}. These states are indeed unlikely to be reached in a platform with a single CPU frequency island, such as the AGX Xavier.
Additionally, while our approach achieves excellent results without resorting to counter multiplexing, optimized event grouping techniques can enable more flexible parameter selection, potentially resulting in more accurate models~\cite{liu2024efficient}. However, time-sharing inherently introduces \gls{pmc} tracking overhead and interpolation errors, requiring a careful trade-off assessment, particularly in real-time applications.}

As far as our contribution to the Linux kernel is concerned, our results pave the way to bring the benefits of our modeling approach to the Linux real-time task scheduler \ccode{SCHED\_DEADLINE} and the CPU frequency governor through the \runmeter framework. This aims to improve the effectiveness and correctness of energy-aware real-time task scheduling within Linux.
Further directions of work also include going beyond the estimation of the current hardware status through predictive models. As of now, the Linux kernel contains very simple linear models for estimating the power consumed at each frequency, which are used to make decisions when selecting the appropriate operating frequency for the CPU.
Models based on online \gls{pmc} data, like that collected by \runmeter, may prove to be more effective from an energy-saving perspective while maintaining a very low overhead.

\section*{Acknowledgement}
This work has received funding from the European Commission through the EU H2020 research project AMPERE (A Model-driven development framework for highly Parallel and EneRgy-Efficient computation supporting multi-criteria optimization) under grant agreement no.\ 871669.

Sergio Mazzola and Gabriele Ara contributed equally to this work.


\bibliography{main.bbl}


\end{document}
\endinput